\documentclass[prb,reprint,aps,superscriptaddress,longbibliography]{revtex4-2}
\usepackage{graphicx}
\usepackage{dcolumn}
\usepackage{bm}
\usepackage{braket}
\usepackage{amsmath}
\usepackage{amssymb}
\usepackage{feynmp}
\usepackage{graphics}
\usepackage{enumerate}
\usepackage[dvipsnames]{xcolor}

\usepackage{comment}

\usepackage[normalem]{ulem} 
\usepackage[hidelinks]{hyperref}

\everymath{\displaystyle}

\newcommand{\bmr}{{\bm r}}

\newcommand{\bmk}{{\bm k}}

\begin{document}

\title{Band mixing and particle-hole asymmetry in moiré fractional Chern insulators}

\author{Nicol\'as Morales-Dur\'an}
\email{nmoralesduran@flatironinstitute.org}
\affiliation{Center for Computational Quantum Physics, Flatiron Institute, New York, New York 10010, USA}

\author{Jingtian Shi}
\affiliation{Materials Science Division, Argonne National Laboratory, Lemont, Illinois 60439, USA}

\author{Cristian Voinea}
\affiliation{School of Physics and Astronomy, University of Leeds, Leeds LS2 9JT, United Kingdom}
\affiliation{Center for Computational Quantum Physics, Flatiron Institute, New York, New York 10010, USA}

\author{Pawe\l{} Potasz}
\affiliation{Institute of Physics, Faculty of Physics, Astronomy and Informatics, Nicolaus Copernicus University, Grudziadzka 5, 87-100 Toru\'n, Poland}

\author{Jennifer Cano}
\affiliation{Department of Physics and Astronomy, Stony Brook University, Stony Brook, New York 11794, USA}
\affiliation{Center for Computational Quantum Physics, Flatiron Institute, New York, New York 10010, USA}

\date{\today}

\begin{abstract}
We investigate the effect of remote band mixing on the stability of fractional Chern insulators in a family of models that approximate continuum descriptions of moiré materials. Our results suggest that the experimentally observed asymmetry between filling fractions $\nu=1/3$ and $\nu=2/3$ in twisted MoTe$_2$ originates from a competition between a fractional Chern insulator, an electron Wigner crystal, and a hole Wigner crystal. In the absence of band mixing, the leading instability at $\nu = 1/3$ is the electron crystal, whereas at $\nu = 2/3$ the main competing phase is the hole crystal. Remote band mixing substantially lowers the energy of the electron crystal but has only a weak effect on the hole crystal. Consequently, it destabilizes the fractional Chern insulator at $\nu=1/3$ more strongly than at $\nu=2/3$. This mechanism also provides an explanation for the emergence of re-entrant integer quantum anomalous Hall states in moiré MoTe$_2$ for fillings $\nu>1/2$.
\end{abstract}

\maketitle

\section{Introduction}

The phase diagram of moir\'e twisted MoTe$_2$ (tMoTe$_2$) has been characterized through a combination of transport, optical and orbital magnetization measurements \cite{cai2023signatures, zeng2023thermodynamic, park2023observation, xu2023observation, Cornell_FQSH1,Redekop2024,Ji2024,Anderson2024trion,Wang2025,xu2025superconductivity,Park2025Ferromagnetism,Xu2025Interplay,Thompson2025Microscopic,Deng2025Nonmonotonic,Li2025Universal,Pan2026optical,Huber2026optical,Park2026observation,Holtzmann2026optical,Park2026,Li2026Signatures,Chen2026visualizing,Cornell_FQSH2,wang2026magnetic,sun2026twistangle}, revealing an extraordinary diversity of correlated electronic states as a function of electron density, displacement field and twist angle. Most strikingly, fractionally filling the topmost moir\'e valence band of tMoTe$_2$ led to the first observation of fractional quantum anomalous Hall (FQAH) states, also known as fractional Chern insulators (FCIs) \cite{Wen_FCI,DasSarma_Sun_FCI,Neupert_FCI,Sheng_FCI,Bernevig_Regnault_FCI}. 
Alongside these developments, re-entrant integer quantum anomalous Hall (RIQAH) states, anomalous Hall metals and signatures of unconventional superconductivity have also been observed  \cite{xu2025superconductivity}. 

Although the FQAH phases are expected to be adiabatically connected to their counterparts in the lowest Landau level (LLL), experiments in tMoTe$_2$ reveal striking departures from the behavior of conventional two-dimensional electron gases in strong magnetic fields. Notably, the correlated topological phases in tMoTe$_2$ are strongest at filling fractions $\nu > 1/2$, with the $\nu = 2/3$ FQAH state emerging as the most prominent, followed by the state at $\nu = 3/5$. In contrast, the $\nu = 1/3$ state -- typically the most robust in the LLL -- was observed only very recently \cite{Pan2026optical}, and earlier experiments instead reported a trivial correlated insulator at the same filling fraction.
These differences raise fundamental questions about the origin and stability of correlated phases in moiré systems.

A substantial body of recent theoretical and numerical work has explored the nature of FQAH ground states in moiré materials and their competition with other correlated phases. Owing to the exponential growth of the Hilbert space, however, most exact diagonalization (ED) \cite{Li2021Spontaneous,Crepel2023Anomalous,Duran2023Pressure,reddy2023fractional,reddy2023toward,Wang2024Fractional} and density matrix renormalization group (DMRG) \cite{dong2023composite,wang2025chiral} studies are restricted to a single isolated band. These studies typically find FCI ground states at both $\nu = 1/3$ and $\nu = 2/3$, albeit with different many-body gaps, consistent with the absence of particle–hole symmetry in bands with a non-uniform quantum metric \cite{Lauchli2013Hierarchy,Abouelkomsan2023Quantum}, and in contrast to the LLL. However, these single-band approaches systematically predict robust FCI states on both sides of $\nu = 1/2$, in disagreement with the experimental observations.

A natural source of this discrepancy is the role of remote bands. Remote band mixing becomes significant when the band gap is comparable to the interaction strength, which is the case across much of the tMoTe$_2$ phase diagram. Multi-band ED studies of continuum moiré models \cite{Bernevig_multibandED,Fu_multibandED,xu2024maximally,goncalves2025spinless} show that including remote bands and the valley degree of freedom destabilizes the FCI state at $\nu = 1/3$, favoring the relative robustness of the $\nu = 2/3$ phase, in closer agreement with experiment. This picture is further supported by recent large-scale approaches, including DMRG \cite{he2025fractional}, neural quantum state variational Monte Carlo (VMC) \cite{li2025deep,luo2025solving}, and similarity renormalization group methods \cite{hou2025stabilizing}, which incorporate band mixing and access larger system sizes. These studies consistently find that while band mixing weakens FCI states in favor of competing charge density wave (CDW) order at both fillings, the FCI remains significantly more stable at $\nu = 2/3$ than at $\nu = 1/3$. Despite this growing numerical consensus, the microscopic mechanism by which band mixing preferentially stabilizes FQAH phases at $\nu > 1/2$ in semiconductor moiré systems such as tMoTe$_2$ remains unclear.

In this work we propose a mechanism for this observation: remote band mixing preferentially stabilizes electron Wigner crystals over hole Wigner crystals. We reach this conclusion by investigating the many-body ground states and low-lying spectra of a family of Hamiltonians that become progressively better approximations to continuum descriptions of semiconductor moiré materials \cite{Duran2024Magic,Shi2024Adiabatic}. Performing ED, we systematically examine the impact of the first remote band on the ground state at fillings $\nu = 1/3$ and $\nu = 2/3$, across three models: Landau levels, Aharonov–Casher (AC) bands \cite{aharonov1979ground}, and adiabatic bands \cite{Duran2024Magic,Shi2024Adiabatic}. This unified approach allows us to disentangle two key sources of particle–hole asymmetry about $\nu = 1/2$: the non-uniform Berry curvature of the active band and the effects of remote band mixing. The response of the many-body gap to LL-mixing differs qualitatively between the two fillings: while the FCI gap at $\nu = 1/3$ decreases monotonically with increasing band mixing, the gap at $\nu = 2/3$ exhibits a non-monotonic evolution, initially increasing before eventually declining. We trace this behavior to a shift in the wave vector of the magnetoroton minimum, which manifests as a level crossing between low-lying excited states in the many-body spectrum. We argue that this level crossing reflects a competition between electron and hole Wigner crystal instabilities, providing a natural explanation for the enhanced robustness of the $\nu = 2/3$ state across a broad parameter regime. We further show that this mechanism remains operative in the presence of non-uniform Berry curvature and finite band width. Finally, we connect our results to prior numerical studies of continuum descriptions of tMoTe$_2$, offering a microscopic explanation for the pronounced particle–hole asymmetry observed experimentally.

\section{Many-body Hamiltonians}
\begin{figure}
    \centering
    \includegraphics[width=0.48\textwidth]{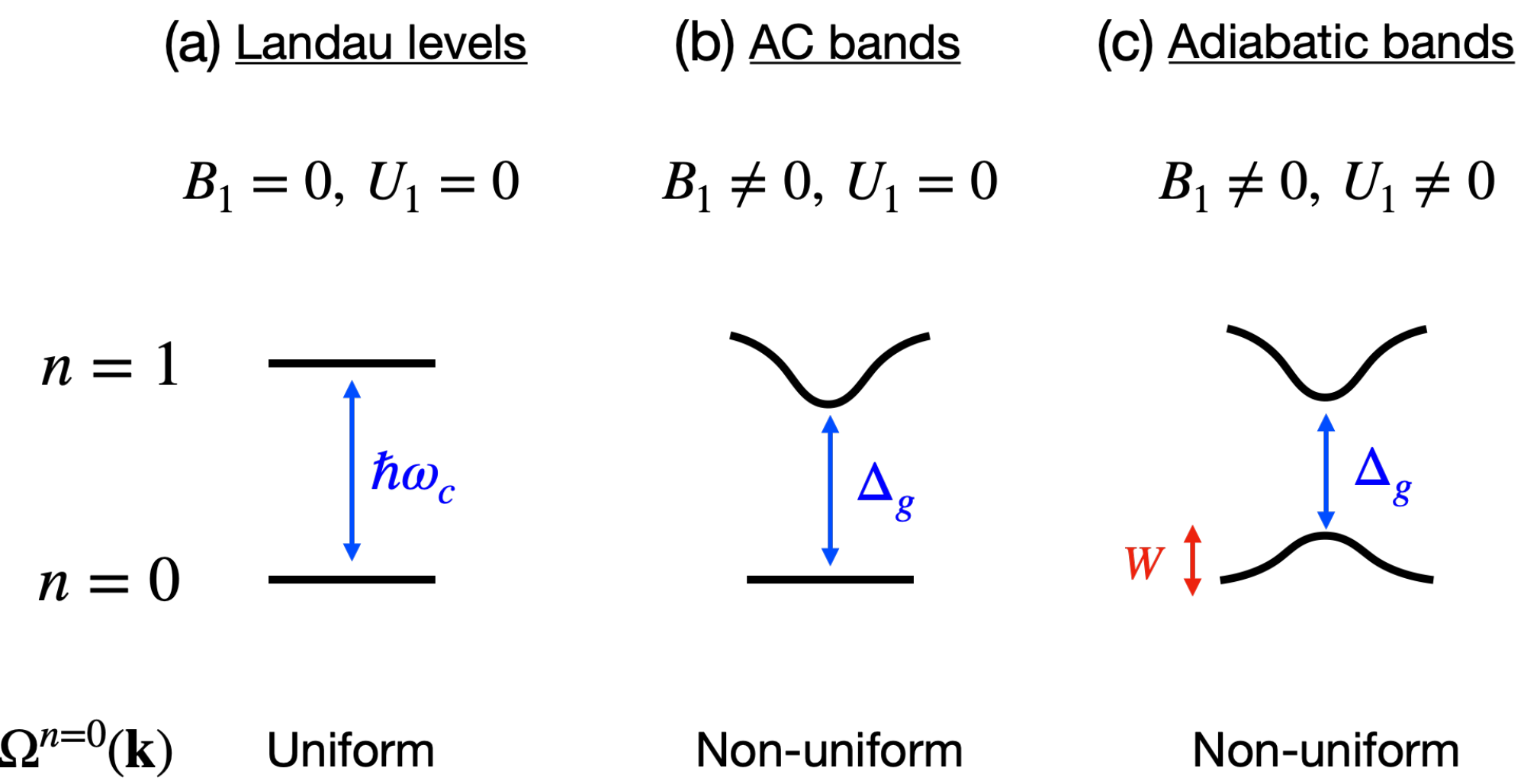}
    
    \caption{Schematics of the different single-particle models considered in this work.
    (a) In Landau levels ($B_1=U_1=0$), all bands are flat with uniform quantum geometry, which is ideal in the LLL.
    (b) In the AC model ($B_1 \neq 0, U_1=0$), the lowest band is flat and has (non-uniform) ideal quantum geometry, while the first remote band is dispersive. (c) In the adiabatic model ($B_1, U_1\neq 0$), all bands are dispersive and the quantum geometry is not ideal. We project interactions to the $n=0$ and $n=1$ bands and obtain the ground state by the ED method. We denote the gap between the two non-interacting bands as $\Delta_g$, which for Landau levels is just the cyclotron gap $\hbar\omega_c$; the band width of the $n=0$ adiabatic band is $W$ and $\Omega^{n=0}$ is the Berry curvature of the lowest band.}
    \label{fig:Schematics}
\end{figure}

We now describe the Hamiltonian for all three models studied here. The single-particle Hamiltonian is that of a Schr\"odinger particle subject to a periodic magnetic field $B({\bm r})$ and a periodic potential $U({\bm r})$, namely
\begin{align}
    H_0=\frac{1}{2m} \Pi_-\Pi_++U({\bm r}),
    \label{eq:Single_Particle_Hamiltonian}
\end{align}
with kinetic-momentum operators given by
\begin{align}
    \Pi_{\pm}= [p_x+A_x({\bm r})]\pm i[p_y+A_y({\bm r})],
\end{align}
where $p_\alpha = -i\partial_\alpha$ for $\alpha = x, y$. The magnetic field can be split into its average and periodic parts,
\begin{align}
    B({\bm r})=\nabla \times {\bm A}(\bmr)=B_0+\delta B({\bm r}).
\end{align}
We take $B({\bm r})$ and $U({\bm r})$ to form triangular lattices with lattice constant $a_M$, and adopt the convention that the magnetic field has one flux quantum per unit cell, which describes the continuum model of tMoTe$_2$ \cite{Duran2024Magic,Shi2024Adiabatic}. This relates $a_M$ to the magnetic length of the average magnetic field, $\ell^2=\hbar/eB_0$, by $\ell^2 =(\sqrt{3}/4\pi)\,a^2_M$. To be concrete, we truncate the periodic functions at their first harmonic (the first-star approximation),
\begin{align}
    \delta B({\bm r})=\sum_{{\bm G}}B_{1}\,e^{i {\bm G}\cdot {\bm r}}, \quad\text{and}\quad U({\bm r})=\sum_{{\bm G}}U_{1}\,e^{i {\bm G}\cdot {\bm r}},
    \label{eq:Bharmonics}
\end{align}
where $\bm G$ are the reciprocal lattice vectors of smallest magnitude. We take $B_1$ to have units of $\ell^{-2}$; $U_1$ has units of energy, $e^2/\varepsilon \ell$, where $\varepsilon$ is the dielectric constant. 
The different models we study are obtained by choosing different values for $B_1$ and $U_1$: the Landau level (LL) Hamiltonian has $B_1=U_1=0$; the AC Hamiltonian has $B_1\neq 0, U_1=0$; and the adiabatic Hamiltonian has both $B_1$ and $U_1$ non-zero; this distinction and consequences for the band structure and quantum geometry are summarized in Fig \ref{fig:Schematics}. Further details on how the adiabatic Hamiltonian in Eq.~\eqref{eq:Single_Particle_Hamiltonian} approximates the continuum model for twisted homobilayer TMDs can be found in \cite{Shi2024Adiabatic,Shi2025effects}.

Electronic interactions are projected into the two lowest bands of Eq. \eqref{eq:Single_Particle_Hamiltonian}, so that the many-body Hamiltonian is
\begin{align}
    H&=\sum_{\bm k, n}\epsilon_{\bm k}\,c^{\dagger}_{\bm k,n}c_{\bm k,n}\nonumber\\
    &+\frac{1}{2}\sum_{i,j,k,l} V_{{\bm k}_i{\bm k}_j{\bm k}_k{\bm k}_l}^{n_in_jn_kn_l}~ c^{\dagger}_{n_i,{\bm k}_i}c^{\dagger}_{n_j,{\bm k}_j}c_{n_l,{\bm k}_l} c_{n_k,{\bm k}_k} ,
    \label{eq:Many_Body_Hamiltonian}
\end{align}
where $\epsilon_{n,{\bm k}}$ is a single-particle energy, $c^{\dagger}_{n, {\bm k}}$ creates a particle in band $n$ with momentum $\bm k$ and the interaction matrix elements are
\begin{align}
    V_{{\bm k}_i{\bm k}_j{\bm k}_k{\bm k}_l}^{n_in_jn_kn_l}=\braket{n_i,{\bm k}_i\,;n_j,{\bm k}_j|V|n_k,{\bm k}_k\,;n_l,{\bm k}_l},
\end{align}
where $\ket{n,{\bm k}}$ are the Bloch states of Eq. \eqref{eq:Single_Particle_Hamiltonian}. Motivated by recent experiments \cite{Pan2026optical} that observe a valley polarized ground state at both $\nu=1/3$ and $\nu=2/3$ we assume spin (valley) polarization and obtain the many-body ground state of the interacting Hamiltonian in Eq. \eqref{eq:Many_Body_Hamiltonian} using ED. For the calculations presented in the main text, we will use the unscreened Coulomb interaction $V({\bm q})=2\pi e^2/\varepsilon{\bm q}$. In Appendix \ref{Appendix:ED_Results} we present additional results for short-ranged interactions corresponding to the first Haldane pseudopotential, $V({\bm q})=-V_1\,{\bm q}^2$. 

The many-body spectrum is determined by three energy scales: the interaction strength $e^2/\varepsilon\ell$; the gap $\Delta_g$ between the two bands in the Hilbert space; and the band width $W$ of the lowest band. 
The band mixing is largely controlled by the ratio of the first two energy scales, $\kappa\equiv(e^2/\varepsilon \ell)/\Delta_g$. To isolate the effect of band mixing on the ground state we have fixed the band gap $\Delta_g$ and tuned the dielectric constant $\varepsilon$. 
We study the effect of $\kappa$ on the many-body gap; 
since the FCI ground state is characterized by a three-fold quasi-degeneracy on the finite torus \cite{Bernevig_Regnault_FCI}, we define the many-body gap as $\Delta=\varepsilon_4-\varepsilon_3$, where $\varepsilon_n$ is the energy of the $n-$th many-body eigenstate. 

The calculations presented in the main text were performed on a system with $N=12$ moiré unit cells and periodic boundary conditions corresponding to the Brillouin zone discretization shown in Fig. \ref{fig:LLL_MBGap}(a). In Appendix \ref{Appendix:ED_Results} we present additional results for other system sizes and boundary conditions, as well as for the sphere geometry. Our results are consistent across all geometries and system sizes considered. 

\section{Landau levels}
We start by taking $B_1=U_1=0$ in Eq. \eqref{eq:Single_Particle_Hamiltonian}, which corresponds to the $n=0$ and $n=1$ LLs. The quantum geometry of the bands is uniform and satisfies $\text{Tr}\,g_{\bm k}=(2n+1)\,\Omega_{\bm k}$, where $g_{\bmk}$ is the quantum metric, $\Omega_{\bm k}$ is the Berry curvature and $n$ is the LL-index. The many-body physics is controlled entirely by the ratio between the interaction energy and the cyclotron gap, namely $\kappa=(e^2/\varepsilon \ell)/(\hbar\omega_c)$. LL-mixing plays a central role in the first ($n=1$) LL, where it strongly influences the competition between the Pfaffian and anti-Pfaffian states at half-filling \cite{Inti_PerturbationTheory,Simon_PerturbationTheory,Nayak_PerturbationTheory1,Nayak_PerturbationTheory2,Shankar_Murthy_Perturbation,Rezayi_antiPfaffian,Wojs_Pfaffian}. 
In contrast, in the LLL ($n=0$), LL-mixing has a more limited impact. For Abelian fractional quantum Hall states such as those at fillings $\nu = 1/3$ and $\nu = 2/3$, 
it does not change the ground state (indeed, these states remain robust even at $\kappa\gg 1$ ~\cite{Yoshioka_1,Yoshioka_2,Yoshioka_3,Jain_crystals}) but LL-mixing can renormalize the gap and energy spectrum.

\begin{figure}
    \centering
    \includegraphics[width=0.48\textwidth]{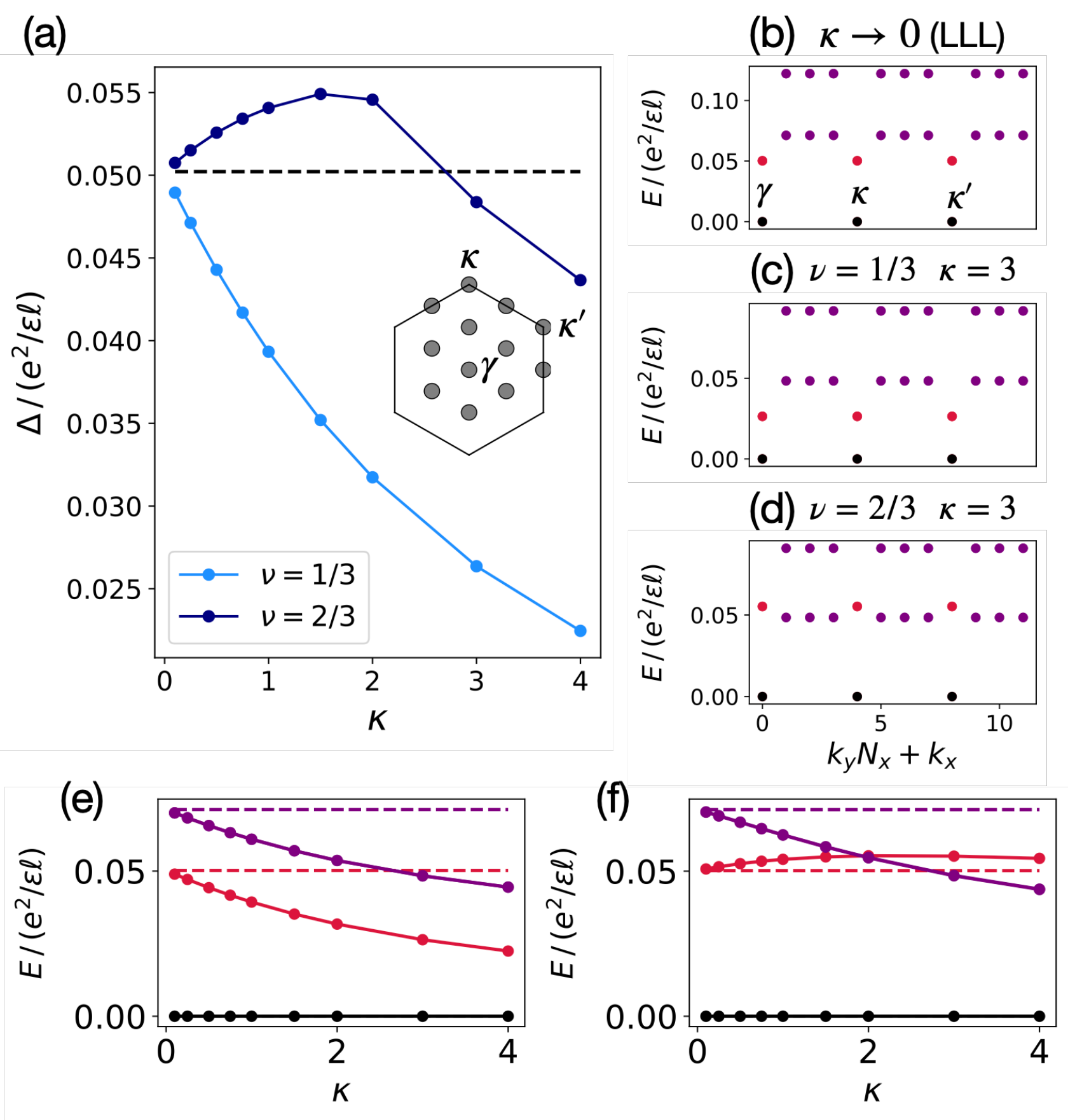}
    
    \caption{Exact diagonalization results for Landau levels. (a) One-band-projected (dashed line) and two-band-projected (lines with dots) many-body gaps for fillings $\nu=1/3,~2/3$ as a function of LL-mixing. Inset indicates the geometry used for ED calculations. (b) Many-body spectrum for LLL-projected calculations; particle-hole symmetry requires the spectra for $\nu=1/3$ and $\nu=2/3$ be identical. (c)-(d) Many-body spectra for two-band-projected calculations with $\kappa=3$ at $\nu=1/3$ and $\nu=2/3$, respectively. (e)-(f) Evolution of the low-energy states as a function of $\kappa$ for filling fractions $\nu=1/3$ and $\nu=2/3$, respectively. Dashed lines correspond to single-band-projected calculations and solid lines to two-band-projected calculations. The states in red correspond to the ${\bm \gamma}, {\bm \kappa}, {\bm \kappa'}-$points and the states in purple to the other nine momentum sectors. At $\nu = 2/3$, the two lowest excitations cross, giving rise to the non-monotonic behavior in the many-body gap in (a).}
    \label{fig:LLL_MBGap}
\end{figure}

Our ED calculations systematically track the behavior of the many-body gap at filling fractions $\nu = 1/3$ and $2/3$ as a function of LL-mixing, revealing trends and differences between the two fillings that extend to the lattice systems we study in subsequent sections.
In Fig. \ref{fig:LLL_MBGap}(a) we show the one-band-projected and two-band-projected many-body gaps at $\nu=1/3$ and $\nu=2/3$, as a function of $\kappa$. The two gaps coincide for calculations projected to the $n=0$ band, as required by the particle-hole symmetry of the LLL. They also coincide in the limit $\kappa\to 0$, which effectively corresponds to a LLL-projection and whose spectrum is shown in Fig.~\ref{fig:LLL_MBGap}(b). However, generically, the two-band-projected many-body gaps are different at the two fillings, and, furthermore, exhibit different trends: while at $\nu=1/3$ the gap decreases monotonically with with LL-mixing \cite{Yoshioka_1}, the gap at $\nu=2/3$ displays non-monotonic behavior as a function of $\kappa$. 
This change in the magnitude of the gap is accompanied by a change in the many-body spectrum: while the fractional quantum Hall ground state survives over the entire range of $\kappa$ considered, strong LL-mixing qualitatively alters the excitation spectrum at $\nu=2/3$, shifting the lowest energy excitation to a different momentum sector. To demonstrate that this is not a finite-size or geometry-dependent effect, in Appendix \ref{Appendix:ED_Results} we show ED results for different torus geometries with $N=12$ and $N=15$ unit cells, along with ED calculations on the sphere geometry. We consistently recover the non-monotonic behavior for the many-body gap at $\nu=2/3$.

We now discuss the many-body spectra in detail.
Typical many-body ED spectra at finite $\kappa$ are shown in Fig. \ref{fig:LLL_MBGap}(c) at $\nu = 1/3$ and (d) at $\nu = 2/3$. Both display three-fold quasi-degenerate ground states at ${\bm \gamma}, {\bm \kappa}, {\bm \kappa'}$, along with many excited states. For two-band calculations, the lowest excited states at $\nu=1/3$ are also at the ${\bm \gamma}, {\bm \kappa}, {\bm \kappa'}-$points, for all values of $\kappa$. In contrast, at $\nu=2/3$ the lowest excited states shift from ${\bm \gamma}, {\bm \kappa}, {\bm \kappa'}$ to the other nine momentum points at a critical value of $\kappa$, whose specific value depends on system size and geometry. In Figs. \ref{fig:LLL_MBGap} (e),(f) we show the evolution of the low-lying many-body energy levels with respect to the ground state as a function of $\kappa$, for filling fractions $\nu=1/3$ and $\nu=2/3$, respectively. The non-monotonic behavior of the many-body gap at $\nu=2/3$ is due to a level-crossing between two excited states, which as we will discuss, reflects a shift in the wave vector of the magnetoroton minimum. For completeness, in Appendix \ref{Appendix:ED_Results} we also present ED results in the limit where the $n=0$ and $n=1$ LLs are degenerate, which applies to the zeroth LL of bilayer graphene. Our results coincide with previous studies \cite{Papic_Abanin_BLG,Jolicoeur_BLG} showing that the gap at $\nu=2/3$ is larger than the gap at $\nu=1/3$, in agreement with experimental observations \cite{BLG_FQH1,BLG_FQH2,BLG_FQH3}. 

\begin{figure}
    \centering
    \includegraphics[width=0.45\textwidth]{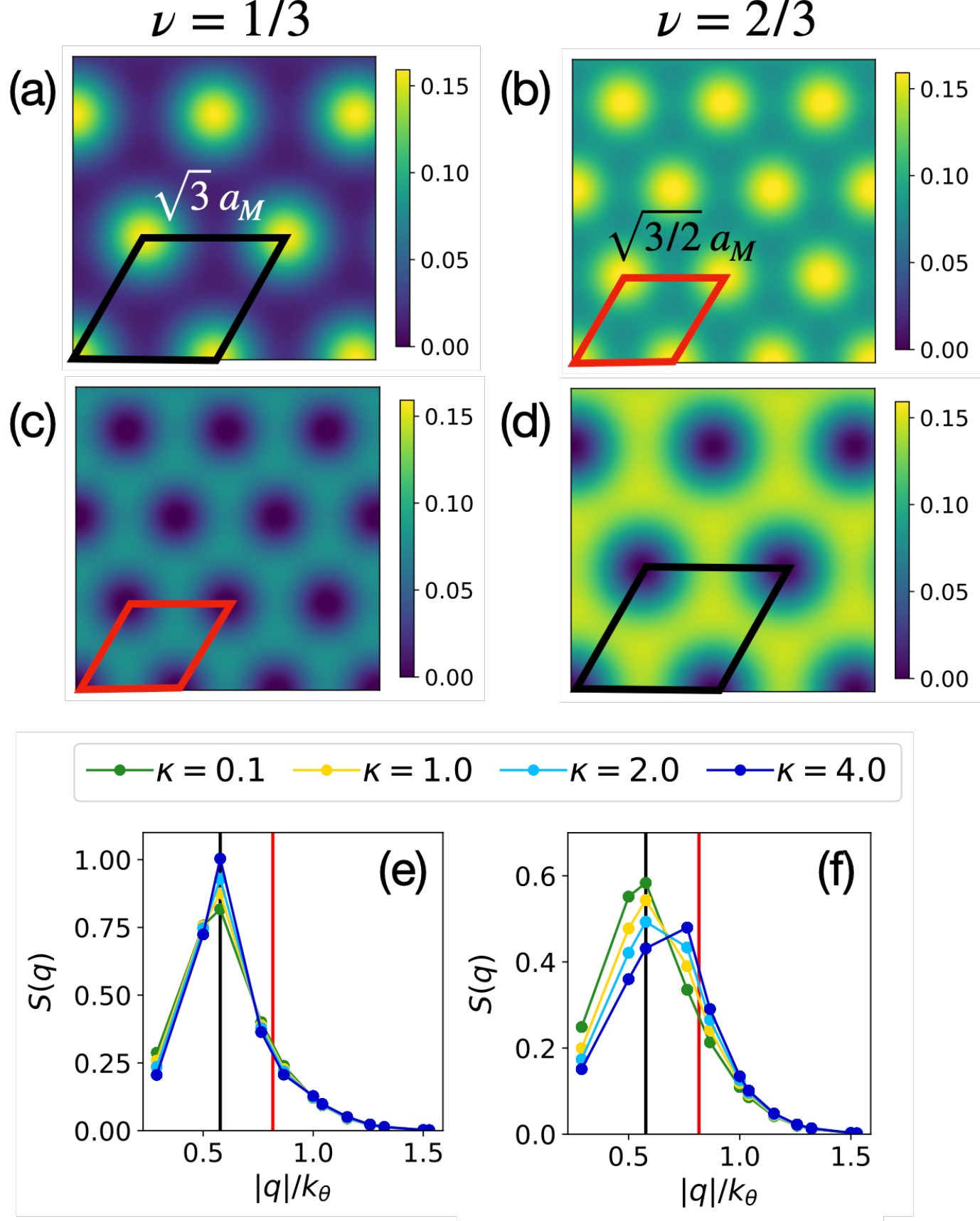}
    
    \caption{Charge densities of the Wigner crystals competing with the fractional quantum Hall states. (a)-(b) Electron crystals at $\nu=1/3$ and $\nu=2/3$, respectively. (c)-(d) Hole crystals at $\nu=1/3$ and $\nu=2/3$, respectively. The unit cells and lattice constants are indicated. The periodicity $a_c=\sqrt{3}~a_M$ is commensurate with that of $B({\bm r})$. For $\nu=1/3$ the electron crystal has lower energy, while for $\nu=2/3$ the hole crystal has lower energy \cite{Macdonald1985Broken}. (e)-(f) Structure factor of the two-band projected FQH ground state for several values of $\kappa$. The vertical lines indicate the wave vectors corresponding to the phase with periodicity $\sqrt{3}\,a_M$ (black) and $\sqrt{3/2}\,a_M$ (red). Momentum is measured in units of $k_{\theta}=4\pi/\sqrt{3}a_M$.}
    \label{fig:LLL_CDW}
\end{figure}

The lowest energy excited state in the ED spectra -- the magnetoroton minimum \cite{Repellin_MRM} -- occurs at a wave vector set by the lattice constant of a competing CDW state. In the context of LLs, the lowest-energy CDW state at $\nu=1/3$ is an electron Wigner crystal, while the most competitive CDW at $\nu=2/3$ is a hole Wigner crystal \cite{Macdonald1985Broken}. Both states have a periodicity $a_c=(4\pi\sqrt{3})^{1/2}~\ell=\sqrt{3}~a_M$, corresponding to a wave vector that coincides with the $\bm \kappa, \bm \kappa'-$ points of the Brillouin zone in our calculations. The electron crystal is topologically trivial, with Chern number $C=0$, while the hole crystal has $C=1$ and is an example of a partial Hall crystal \cite{Tesanovic_HallCrystal,Huang_2025_Apparent}.
Other competing states (which are higher in energy in the absence of band mixing~\cite{Macdonald1985Broken}) are the hole crystal at $\nu=1/3$ and the electron crystal at $\nu=2/3$.
These two states share the periodicity $a_c=\sqrt{3/2}~a_M$, which is incommensurate with the periodicity of the fields $B({\bm r})$ and $U({\bm r})$ included in later sections; consequently, their reciprocal lattice vectors do not coincide with any of the Brillouin zone points in our calculations. The charge density in all four states is schematically shown in Fig.~\ref{fig:LLL_CDW}(a)-(d). 

In Fig. \ref{fig:LLL_CDW}(e),(f) we show the FQH ground state structure factor $S({\bm q})=\braket{\rho_{\bm q}\rho_{-{\bm q}}}/N_e$, where $N_e$ is the number of particles and $\rho_{\bm q}$ is the band-projected density \cite{Zaklama_StructureFactor}, for different values of $\kappa$ at $\nu=1/3$ and $\nu=2/3$. We observe a peak at the $\bm \kappa,\bm \kappa'-$points for $\nu=1/3$, which slightly grows in magnitude with LL-mixing. In contrast, for $\nu=2/3$ the main peak in the structure factor is at $\bm \kappa,\bm \kappa'$ for small $\kappa$ but moves to a larger wave vector for large $\kappa$. 
This is consistent with the many-body spectrum on the sphere, where, as a function of angular momentum, the magnetoroton minimum shifts to larger wave vectors with increasing LL mixing (see Appendix \ref{Appendix:ED_Results} for details).

LL-mixing is known to lower the energy of electron Wigner crystals more strongly than that of their hole counterparts \cite{macdonald1984influence, Louie_LLMixing}.
We therefore interpret the level crossing at $\nu = 2/3$ as a signature of a change in the dominant CDW instability: at small $\kappa$, the leading instability is toward a hole Wigner crystal with periodicity $a_c=\sqrt{3}\,a_M$, but as $\kappa$ increases, LL-mixing progressively stabilizes the electron crystal with periodicity $a_c=\sqrt{3/2}\,a_M$. 
This transition naturally explains the observed shift of the magnetoroton minimum to larger wave vectors.

In the following sections, we show that the qualitative dependence of the FQH many-body gaps at $\nu = 1/3$ and $\nu = 2/3$ on LL-mixing persists in both the AC and adiabatic models. 

\section{Aharonov-Casher bands}

We now turn to the AC band, where the magnetic field $B(\bm r)$ in Eq.~\eqref{eq:Single_Particle_Hamiltonian} is spatially modulated, but the potential $U(\bm r)$ is absent. For concreteness, we keep only the first harmonic $B_1$ of the magnetic field, as described in Eq.~\eqref{eq:Bharmonics}. As illustrated in Fig. \ref{fig:Schematics}, the $n=0$ band is flat and ideal, but has non-uniform Berry curvature, while the $n=1$ band is dispersive and its quantum geometry is not equal to that of the $n=1$ LL. Due to the non-uniform Berry curvature, particle-hole symmetry is broken at the one-band-projected level, while due to the remote band dispersion, the band gap $\Delta_g$ is smaller than the cyclotron gap in the limit $B_1=0$. Hence, we define the band-mixing parameter as $\kappa \equiv (e^2/\varepsilon \ell)/\Delta_g$. As we will show, our calculations of the many-body spectrum disentangle the effects of quantum geometry and remote band mixing on the particle-hole asymmetry.

\begin{figure}
    \centering
    \includegraphics[width=0.48\textwidth]{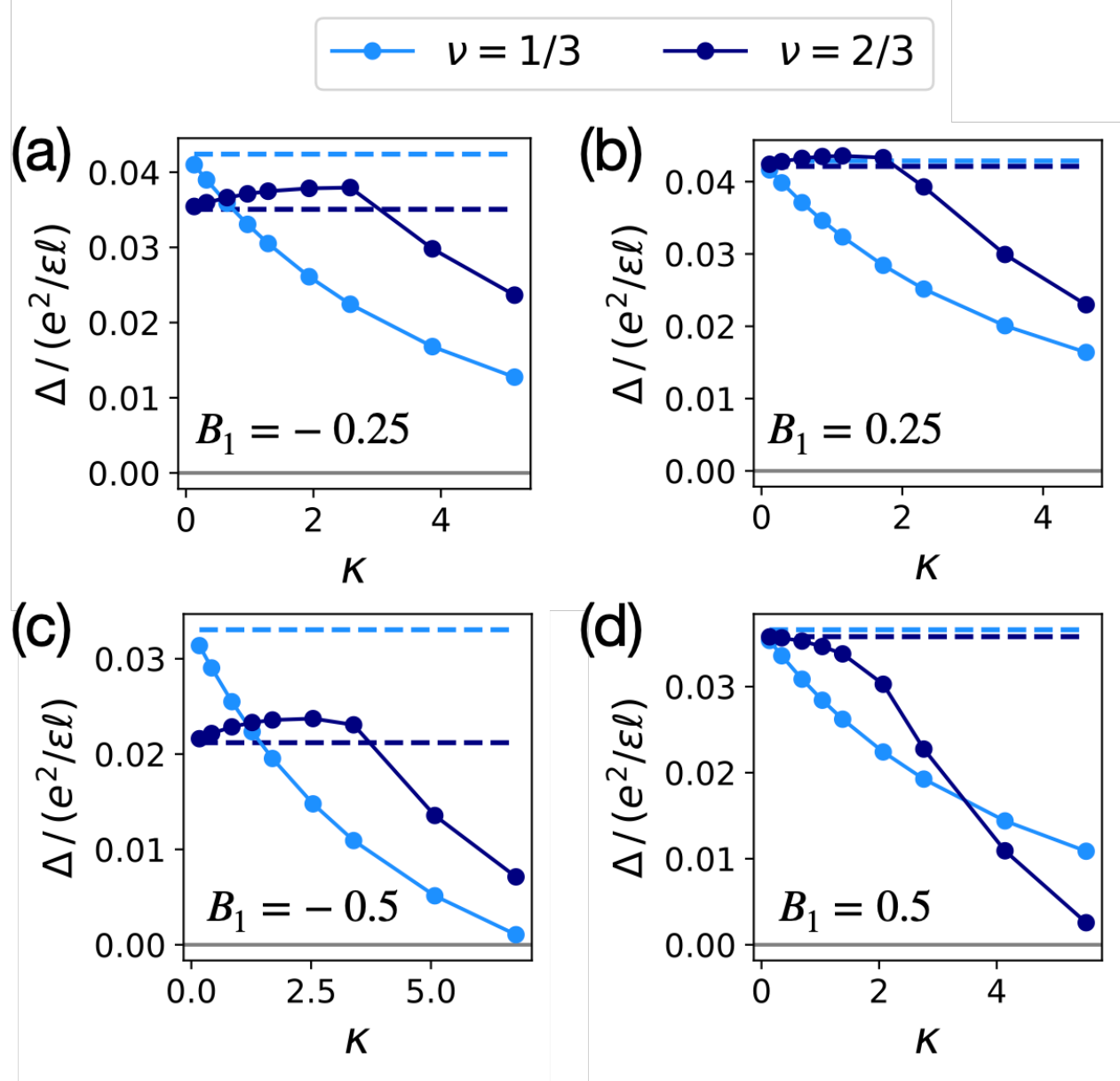}
    
    \caption{Exact diagonalization results for AC bands with different values of $B_1$, as indicated. The Brillouin zone geometry is the same as in Fig. \ref{fig:LLL_MBGap}. One-band-projected (dashed line) and two-band-projected (lines with dots) many-body gaps for fillings $\nu=1/3,~2/3$ as a function of band-mixing. Due to the non-uniform quantum geometry of the $n=0$ band, the gap at $\nu = 1/3$ exceeds that at $\nu = 2/3$ for one-band-projected calculations. Band mixing, however, can reverse this ordering.}
    \label{fig:AC_MBGap}
\end{figure}

Fig.~\ref{fig:AC_MBGap} shows the behavior of the many-body gaps at fillings $\nu=1/3$ and $\nu=2/3$ for AC bands with different values of $B_1$ as a function of the band-mixing parameter $\kappa$. The single-band-projected gaps (dashed lines) remain constant, although they no longer coincide with each other due to the particle-hole asymmetry introduced by the non-uniform quantum geometry \cite{Lauchli2013Hierarchy,Abouelkomsan2023Quantum}. For the $N=12$ geometry considered here, the particle-hole asymmetry is noticeable for $B_1<0$ and almost vanishing for $B_1>0$. In Appendix \ref{Appendix:ED_Results} we show similar ED results for different boundary conditions and system sizes, which indicate that the particle-hole asymmetry coming from the quantum metric of the $n=0$ AC band depends significantly on the momentum discretization for small systems, and that this dependence becomes less severe as the system size is increased. Nonetheless, we consistently observe that the many-body gap at $\nu=1/3$ is larger than at $\nu=2/3$ for all calculations projected onto the $n=0$ AC band. Single-band-projected ED studies of continuum models of tMoTe$_2$ \cite{Li2021Spontaneous,Duran2023Pressure,reddy2023fractional,reddy2023toward,Wang2024Fractional} observe that the two gaps have similar magnitudes, suggesting that the quantum geometry of the topmost band alone is insufficient to explain the experimental particle-hole asymmetry, which favors the FCI at $\nu = 2/3$.

\begin{figure}
    \centering
    \includegraphics[width=0.45\textwidth]{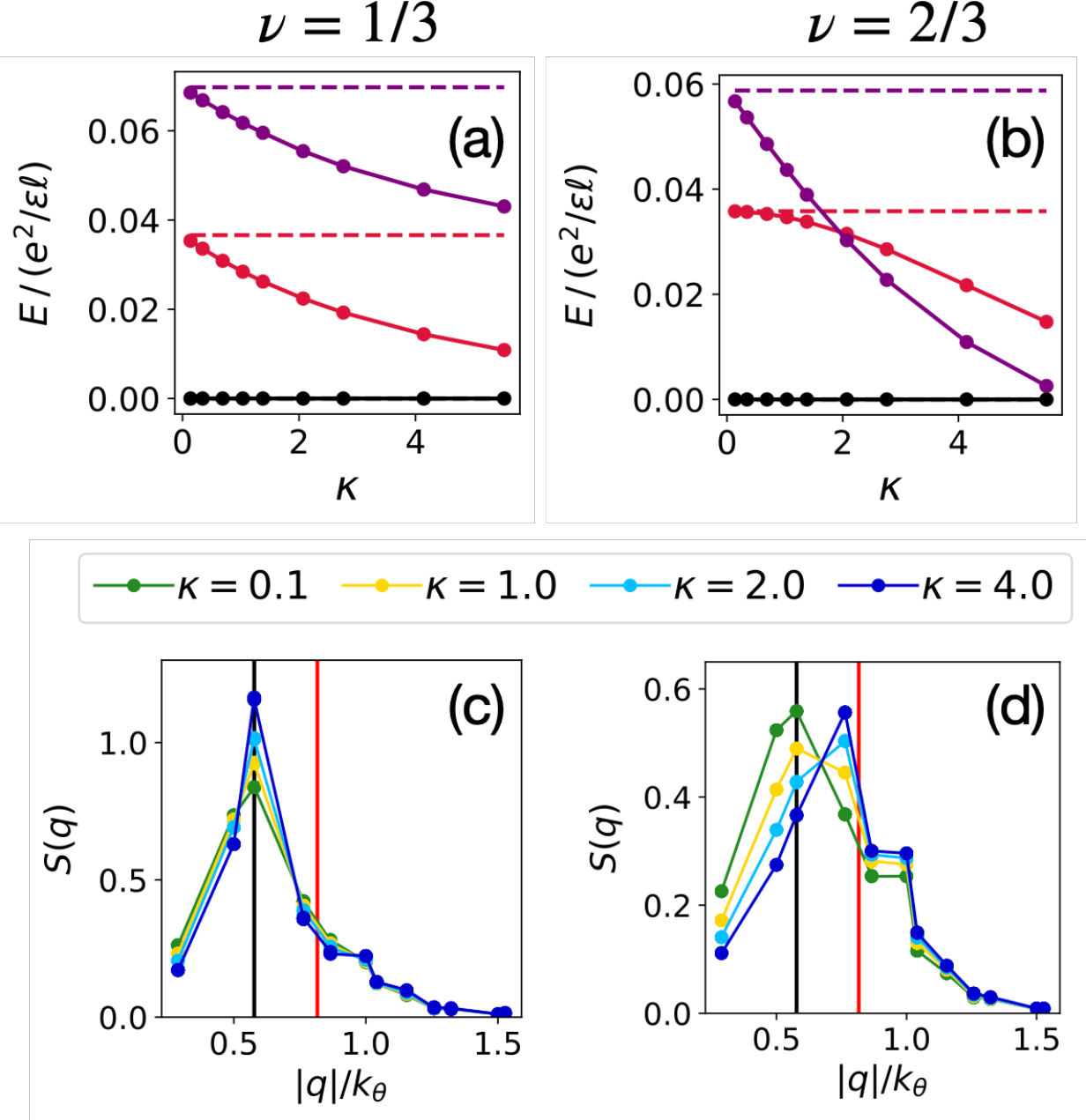}
    
    \caption{Evolution of the low-energy many-body states as a function of $\kappa$ at (a) $\nu=1/3$ and (b) $\nu=2/3$ for an AC band with $B_1=0.5$. Dashed lines correspond to single-band-projected calculations and lines with dots to two-band-projected calculations. The color convention is the same as for Fig. \ref{fig:LLL_MBGap}. Ground state structure factors at (c) $\nu=1/3$ and (d) $\nu=2/3$ for several values of $\kappa$. The vertical lines indicate the wave vectors corresponding to the crystal unit cells shown in Fig. \ref{fig:LLL_CDW}. The level crossing in (b) correlates with the shift in the main peak of $S(q)$ in (d). Note that a secondary peak develops at the first shell of reciprocal lattice vectors $|q|=k_{\theta}$.}
    \label{fig:StructureFactor_AC}
\end{figure}

The effect of band mixing with the $n=1$ AC band induces an additional particle-hole asymmetry above and below $\nu=1/2$, as can be seen from the two-band-projected many-body gaps in Fig. \ref{fig:AC_MBGap} (lines with dots). The gap at $\nu=1/3$ decreases monotonically as a function of $\kappa$ for all values of $B_1$, while the gap at $\nu=2/3$ displays a behavior similar to the one observed for Landau levels. In Fig. \ref{fig:StructureFactor_AC}(a)-(b) we show the evolution of the lowest many-body energy levels with $\kappa$ for an AC band with $B_1=0.5$ and for both filling fractions $\nu=1/3$ and $\nu=2/3$, respectively. There is a level crossing at $\nu=2/3$ which explains why the many-body gap behaves different than its counterpart at $\nu=1/3$. In Fig. \ref{fig:StructureFactor_AC}(c)-(d) we show the ground state structure factor at $\nu=1/3$ and $\nu=2/3$, respectively, for different values of $\kappa$. Similarly to the LL case, at $\nu = 1/3$, the main peak always remains at the ${\bm \kappa}/{\bm \kappa}'-$points, but at $\nu = 2/3$, it shifts to a larger momentum when $\kappa$ is large.
This shift is consistent with the dominant CDW instability shifting from a hole to an electron Wigner crystal. 

\section{Adiabatic bands}
\begin{figure*}
    \centering
    \includegraphics[width=0.95\textwidth]{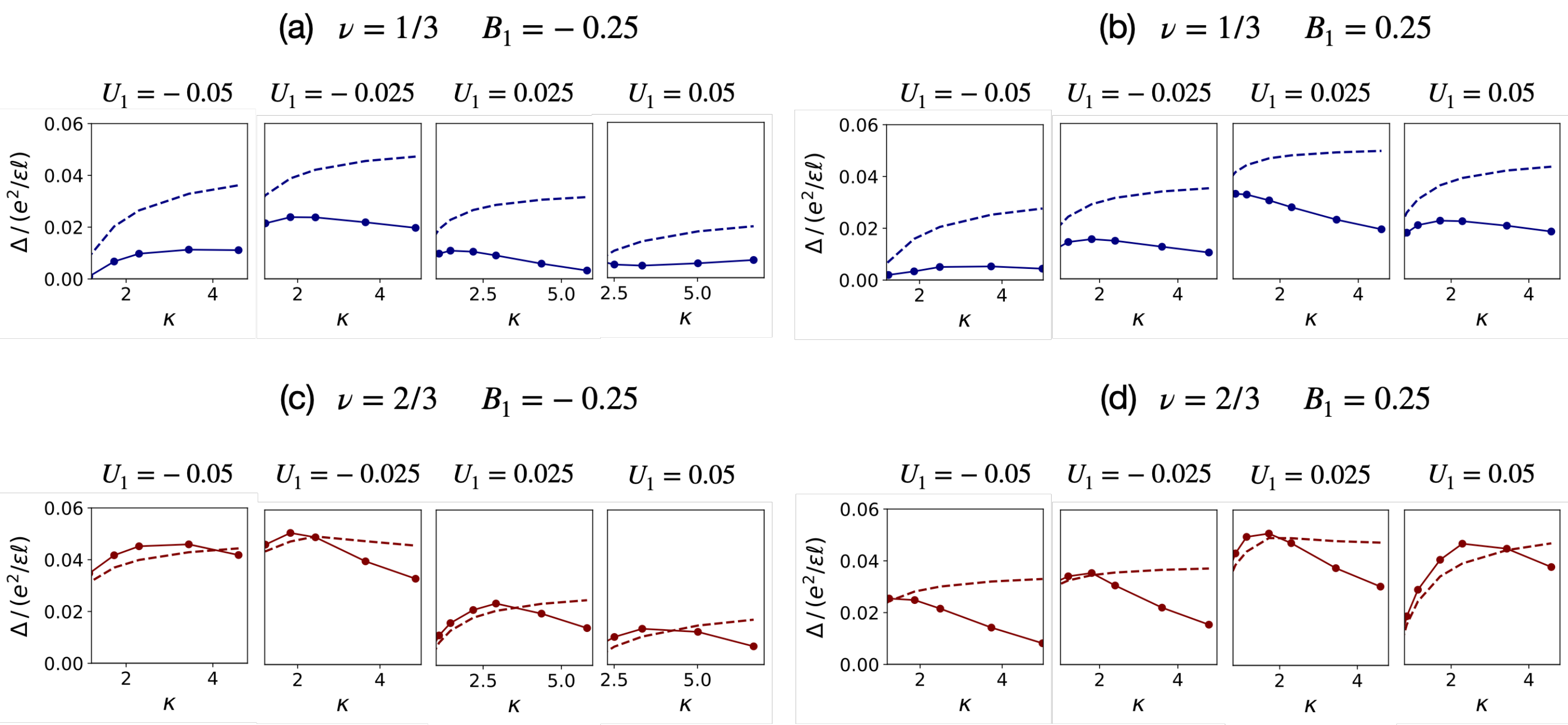}
    
    \caption{Exact diagonalization results for adiabatic bands. Evolution of the many-body gaps as a function of $\kappa$ for different adiabatic bands at filling fractions (a),(b) $\nu=1/3$ and (c),(d) $\nu=2/3$. We have fixed $B_1\neq 0$ in all calculations and vary $U_1$ as indicated in each panel. Dashed lines and lines with dots are one-band and two-band-projected calculations, respectively. The one-band-projected gaps increase with $\kappa$ and asymptote to a constant value in the limit $e^2/\varepsilon\ell\gg W$ (equivalently as $\kappa\to\infty$).}
    \label{fig:Adiabatic_MBGap}
\end{figure*}
\begin{figure}
    \centering
    \includegraphics[width=0.4\textwidth]{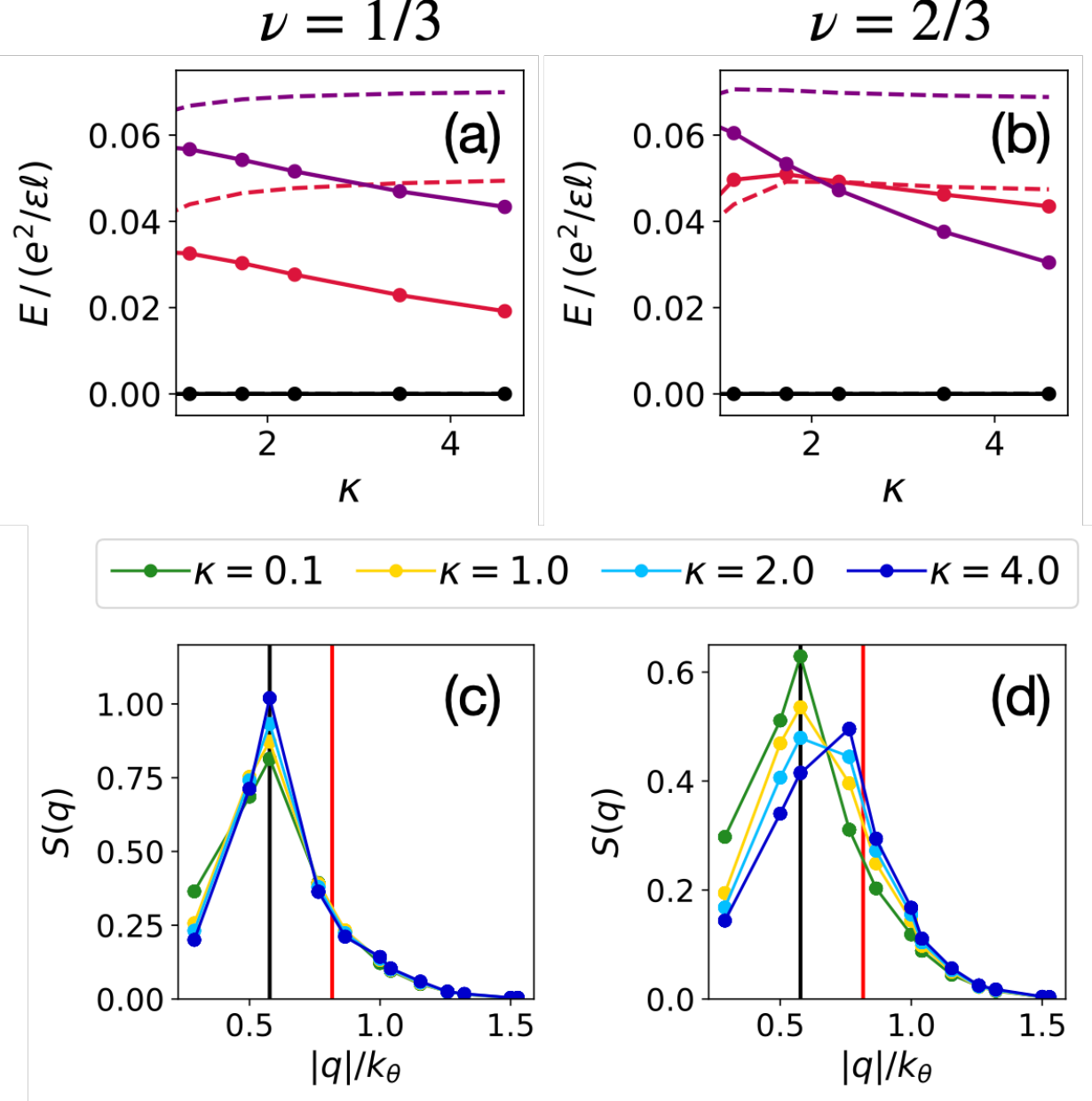}
    
    \caption{Evolution of the low-energy many-body states as a function of $\kappa$ at (a) $\nu=1/3$ and (b) $\nu=2/3$ for an adiabatic band with $B_1=0.25$ and $U_1=0.025$. Dashed lines correspond to single-band-projected calculations and lines with dots to two-band-projected calculations. The color convention is the same as for Fig. \ref{fig:LLL_MBGap}. Ground state structure factors at (c) $\nu=1/3$ and (d) $\nu=2/3$ for several values of $\kappa$. The vertical lines indicate the wave vectors corresponding to the crystal unit cells shown in Fig. \ref{fig:LLL_CDW}. The level crossing in (b) at filling $\nu = 2/3$ correlates with the shift in the main peak of $S(q)$ in (d). Secondary peaks at $|q|=k_{\theta}$ start developing but are less pronounced than in Fig. \ref{fig:StructureFactor_AC}.}
    \label{fig:StructureFactor_Adiabatic}
\end{figure}

By taking both $B_1\neq0$ and $U_1\neq 0$ in Eq.~\eqref{eq:Single_Particle_Hamiltonian}, we obtain a pair of dispersive bands with non-ideal quantum geometry, as indicated in Fig. \ref{fig:Schematics}. The parameter $U_1$ controls the band width of the $n=0$ band, which we denote $W$. As commented above, the many-body physics of the adiabatic model is dictated by three energy scales: $e^2/\varepsilon\ell$, $\Delta_g$ and $W$. For sufficiently large $W$, the two lowest bands can invert, changing the Chern number of the $n=0$ adiabatic band from $C=1$ to $C=0$. 
Since we focus on the FCI ground state, we exclude the topologically trivial regime by restricting to $W\ll \Delta_g$ and tuning the interaction strength via the dielectric constant $\varepsilon$. 
Moreover, because the interacting state of a fractionally filled $n=0$ band is expected to be metallic (with a vanishing many-body gap) when $e^2/\varepsilon\ell \lesssim W$, we further confine ourselves to the regime where interactions dominate over the bandwidth. 

In Fig. \ref{fig:Adiabatic_MBGap} we show the evolution of the many-body gaps at fillings $\nu=1/3$ and $\nu=2/3$, for different adiabatic bands. 
For $\nu=1/3$ the two-band-projected gaps (lines with dots) are consistently smaller than the one-band-projected gaps (dashed lines). Meanwhile, for $\nu=2/3$ we can identify a range of $\kappa$ for which the two-band-projected gap is larger than the one-band-projected gap, which reverses for large enough $\kappa$. This behavior follows our observations for Landau levels and AC bands, and can be explained by a level crossing between excited states, as can be seen in Figs. \ref{fig:StructureFactor_Adiabatic}(a)-(b) for an adiabatic band with $B_1=0.25$ and $U_1=0.025$. Fig. \ref{fig:StructureFactor_Adiabatic}(c)-(d) shows the ground state structure factors at fillings $\nu=1/3$ and $\nu=2/3$ of the same band, which again reveals a shift in the momentum of the main peak of $S({\bm q})$ at $\nu=2/3$. Note also that for all the adiabatic bands considered in Fig. \ref{fig:Adiabatic_MBGap}, the two-band projected gaps at $\nu=2/3$ are larger than those at $\nu=1/3$, meaning that the FCI at $\nu=2/3$ is more robust.

\section{Continuum model for moiré TMDs}
In this section we present one-band and two-band-projected ED results for the continuum model of tMoTe$_2$. Similarly to the case of adiabatic bands, there are three energy scales that determine the many-body physics, namely the interaction scale $e^2/\varepsilon \ell$ (recall $\ell^2=(\sqrt{3}/4\pi)a_M^2$), the band gap $\Delta_g$ and the band width of the topmost band $W$. We have chosen two sets of continuum model parameters. The first model \cite{reddy2023fractional} corresponds to parameters $(V,\psi,\omega)=(11.2,-91^{\circ}, -13.3)$ and twist angle $\theta=2.0^{\circ}$. For this model, the two topmost valence bands have the same Chern number, and a recent VMC study \cite{li2025deep} found the FCI at $\nu=2/3$ to be stable to remote band mixing, while the ground state at $\nu=1/3$ is found to be a CDW. The second model corresponds to parameters $(V,\psi,\omega)=(16.5,-105.9^{\circ}, -18.8)$ \cite{Bernevig_DFT} and $\theta=3.5^{\circ}$, which is approximately the twist angle of the sample studied in \cite{Pan2026optical}. In this model the Chern numbers of the two topmost moiré bands are opposite.

\begin{figure}
    \centering
    \includegraphics[width=0.48\textwidth]{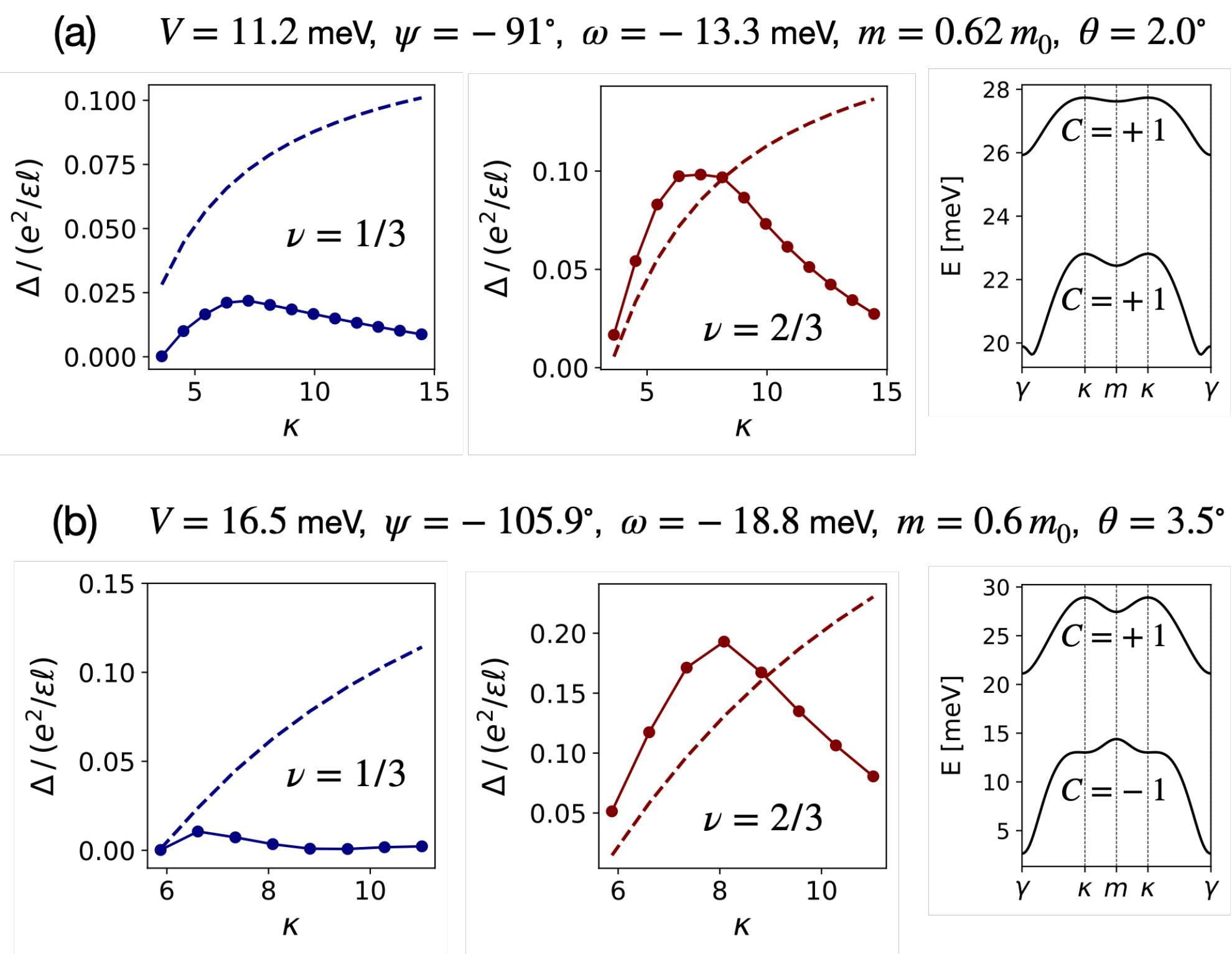}
    
    \caption{Exact diagonalization results for continuum models of tMoTe$_2$. Evolution of the many-body gaps at $\nu=1/3$ and $\nu=2/3$ as a function of band mixing. Dashed lines correspond to single-band-projected calculations and lines with dots to two-band-projected calculations The rightmost panel shows the two topmost $K-$valley non-interacting bands with their respective Chern numbers; the bands emerging from the $K'-$valley are related via time-reversal.}
    \label{fig:Continuum_MBGap}
\end{figure}

In Fig. \ref{fig:Continuum_MBGap} we show the evolution of the many-body gaps at filling fractions $\nu=1/3$ and $\nu=2/3$ with respect to band-mixing for both continuum models. The behavior of the gaps is qualitatively similar to the ones shown in Fig. \ref{fig:Adiabatic_MBGap} for adiabatic bands. The one-band-projected results (dashed lines) are always larger than their two-band-projected counterparts (lines with dots) at $\nu=1/3$. In contrast, at $\nu=2/3$ there is a range of $\kappa$ for which the two-band-projected gap is larger than the one-band-projected gap, which reverses for large enough $\kappa$. This behavior can be associated with a level crossing between excited states, as we show in Appendix \ref{Appendix:ED_Results}. In addition, for a fixed value of $\kappa$, the many-body gap at $\nu=2/3$ is always larger than the gap at $\nu=1/3$, indicating that the FCI states are more stable at filling fractions $\nu>1/2$ when band-mixing in included, which is consistent with previous ED studies \cite{Bernevig_multibandED,Fu_multibandED}.

A recent multi-band ED study \cite{goncalves2025spinless} of the model \cite{Bernevig_DFT} which considers a slightly larger twist angle $\theta=3.7^{\circ}$ finds that the magnetoroton gap at $\nu = 2/3$ is nearly unchanged with the inclusion of band mixing.
They considered a value of the dielectric constant relevant to experiment, $\varepsilon\approx11.11$, which translates into $\kappa\sim 9$. 
This is consistent with our calculation: Fig. \ref{fig:Continuum_MBGap}(b) shows that the level crossing between excited energy levels happens very near $\kappa=9$, and that the one-band and two-band projected gaps have similar magnitudes at that point in the phase diagram. This explains the robustness of the FCI at $\nu=2/3$ seen in \cite{goncalves2025spinless}. 
This is also consistent with the experimental observation of a RIQH effect with $C=1$ when the density is tuned away from $\nu = 2/3$ \cite{xu2025superconductivity}, since this observation indicates that the main instability is a hole Wigner crystal, which onsets around $\kappa \lesssim 8$ according to our results. 

\section{Discussion}
Previous work \cite{Bernevig_multibandED,Fu_multibandED,goncalves2025spinless,xu2024maximally,he2025fractional,li2025deep,luo2025solving,hou2025stabilizing} has established that incorporating remote bands is essential for correctly reproducing the experimental phenomenology of tMoTe$_2$. 
However, the mechanism driving this effect was unclear.
Motivated by experimental observations of RIQAH states and the particle-hole asymmetry in the phase diagram of tMoTe$_2$, we have investigated the effect of remote band mixing on the nature of the FCI ground states at filling fractions $\nu=1/3$ and $\nu=2/3$. By exploiting the mapping between continuum moiré TMD models and an effective Landau level problem \cite{Duran2024Magic}, we leveraged known results on CDWs in the fractional quantum Hall effect to propose an explanation for the particle–hole asymmetry observed in numerical studies of tMoTe$_2$ continuum Hamiltonians, which may also account for recent experimental findings \cite{Pan2026optical}.

All the models that we studied here, namely Landau levels, AC bands, adiabatic bands and continuum moiré models, exhibit a larger many-body gap at $\nu = 2/3$ than at $\nu = 1/3$ and a two-step behavior in the multi-band projected gap at $\nu=2/3$ as a function of band mixing.
This characteristic behavior of the $\nu=2/3$ gap is due to a crossing between the two lowest excited energy levels, which we attribute to a change in the dominant CDW instability of the FCI state. In the context of Landau levels, LL-mixing allows for electron Wigner crystals to lower their energy significantly, while the energy of the hole crystals remains largely unchanged \cite{macdonald1984influence}. Thus, at filling $\nu=2/3$ and $\kappa\to0$, the hole crystal initially has lower energy than the electron crystal \cite{Macdonald1985Broken}, 
but at a critical value of $\kappa$, the electron crystal becomes the lowest-energy competing CDW phase. This transition is reflected in the many-body spectrum by a change in the position of the magnetoroton minimum, and is accompanied by shift in the wave vector where the FQH structure factor is peaked. The increase of the magnetoroton minimum wave vector is consistent with the unit cell of the main competing CDW becoming smaller. A recent VMC study of LLs on the sphere confirms this picture \cite{zhu2026crystallization}. Going beyond LLs, we have shown that this behavior persists in the AC and adiabatic models and even in some regimes of the continuum model for tMoTe$_2$, although the confluence of finite band width and non-trivial quantum geometry makes it harder to ascribe the particle-hole asymmetry to a single cause. The mechanism described here offers an explanation for why the inclusion of the second band in previous ED studies~\cite{Fu_multibandED,Bernevig_multibandED,xu2024maximally,goncalves2025spinless} weakens the state at $\nu=1/3$ more than at $\nu=2/3$. 

It is also worth noting that the impact of band mixing on the FCI ground state may depend on the Chern number and quantum geometry of the remote band. In nearly all cases considered here, the two bands share the same Chern number and can therefore be viewed as deviations from a generalized Landau level structure. However, in certain experimentally relevant twist-angle regimes of tMoTe$_2$, the two topmost bands carry opposite Chern numbers \cite{Xu2025Interplay,Park2025Ferromagnetism} -- as in the model shown in Fig. \ref{fig:Continuum_MBGap}(b) -- for which a Haldane-like description is more appropriate \cite{he2025fractional}. While such differences in band structure may lead to quantitative modifications, we expect the central mechanism identified here -- namely that remote band mixing preferentially stabilizes electron Wigner crystals over hole Wigner crystals -- to remain robust. This provides a natural explanation for the enhanced stability of the $\nu = 2/3$ FCI and the proliferation of RIQAH states in tMoTe$_2$ when the density is slightly away from $\nu=2/3$. Finally, we note that disorder and strain effects, which are not included in our study but are present in experiments, may further contribute to the observed particle–hole asymmetry.\\

{\it Acknowledgements--} We thank Allan H. MacDonald, Zlatko Papi\'c and Nicolas Regnault for fruitful discussions. This work was performed in part at the Aspen Center for Physics, which is supported by National Science Foundation grant PHY-2210452. J.S. acknowledges support by the US Department of Energy, Office of Science, Basic Energy Sciences, Materials Sciences and Engineering Division. C.V. acknowledges support by the Leverhulme Trust Research Leadership Award RL-2019-015 and EPSRC Grants EP/Z533634/1,
UKRI1337. Statement of compliance with EPSRC policy framework on research data: This publication is theoretical work that does not require supporting research data. PP acknowledges the Texas Advanced Computing Center (TACC) at The University of Texas at Austin for providing high-performance computer resources. The Flatiron Institute is a division of the Simons Foundation.

\bibliography{bibliography}

\clearpage
\appendix
\section{Additional exact diagonalization results}
\label{Appendix:ED_Results}
In this Appendix we present additional ED results on different geometries and system sizes to complement the results reported in the main text.

\subsection{Torus geometry}
\begin{figure*}
    \centering
    \includegraphics[width=\textwidth]{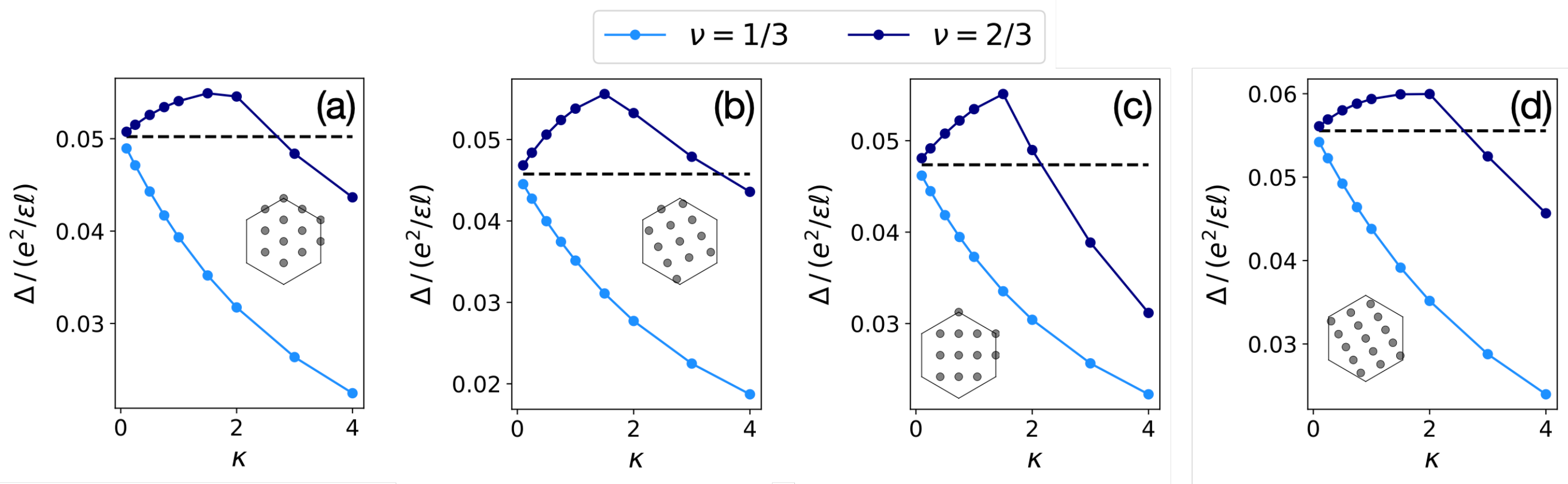}
    
    \caption{Exact diagonalization results on Landau levels for different Brillouin zone discretizations. One-band-projected (dashed line) and two-band-projected (lines with dots) many-body gaps for fillings $\nu= 1/3, 2/3$ as a function of LL-mixing. The insets indicate the geometry used for calculations. (a)-(c) correspond to $N=12$ unit cells and (d) to $N=15$ unit cells. We recover the same contrasting behavior between $\nu=2/3$ and $\nu=1/3$ in all the system sizes and discretizations that we study.}
    \label{fig:LLL_Gap_Geometries}
\end{figure*}

In Fig. \ref{fig:LLL_Gap_Geometries} we show the evolution of the many-body gap at fillings $\nu=1/3$ and $\nu=2/3$ of Landau levels, as a function of $\kappa$. We present results for different choices of periodic boundary conditions, which translate into different discretizations of the Brillouin zone. We observe that the behavior of the gaps at $\nu=1/3$ and $\nu=2/3$ is consistent across all systems studied.

\begin{figure*}
    \centering
    \includegraphics[width=\textwidth]{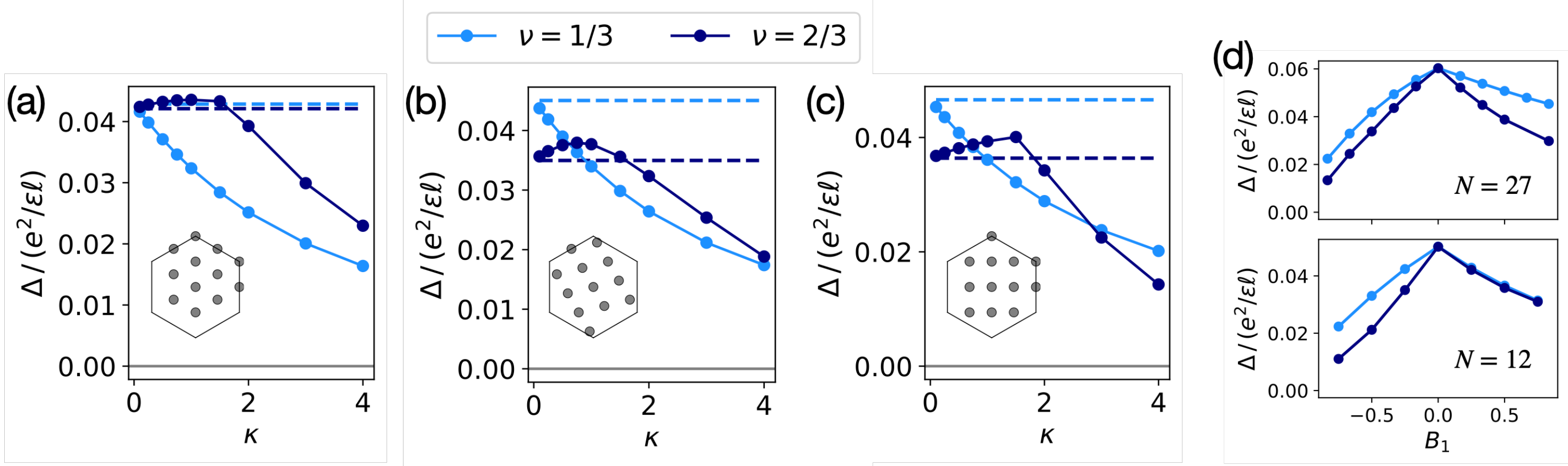}
    
    \caption{Particle-hole asymmetry above and below $\nu=1/2$ in an AC band with $B_1=0.25$. (a)-(c) Evolution of the many-body gap at fillings $\nu=1/3$ and $\nu=2/3$, as a function of $\kappa$ and for different Brillouin zone discretizations which are indicated in the insets. (d) Evolution of the one-band-projected many-body gaps as a function of $B_1$ for two system sizes. The asymmetry with respect to the sign of $B_1$ becomes less pronounced as system size increases. For all the AC models studied here, the gap at $\nu=1/3$ is greater than its counterpart at $\nu=2/3$ at the single-band-projected level.}
    \label{fig:AC_Gap_Geometries}
\end{figure*}

In Fig \ref{fig:AC_Gap_Geometries} we show the evolution of the many-body gap as a function of $\kappa$ for an AC band with $B_1=0.25$ and for different Brillouin zone discretizations in a system with $N=12$ unit cells. As commented in the main text, because the Berry curvature is not uniform, the particle-hole symmetry is already broken at the single-band-projected level, which is manifested in the two dashed lines not coinciding. We see that the amount of particle-hole asymmetry at the single-band-projected level depends on the Brillouin zone discretization. Despite this, the effect of remote bands on particle-hole asymmetry is always consistent.

\begin{figure*}
    \centering
    \includegraphics[width=0.9\textwidth]{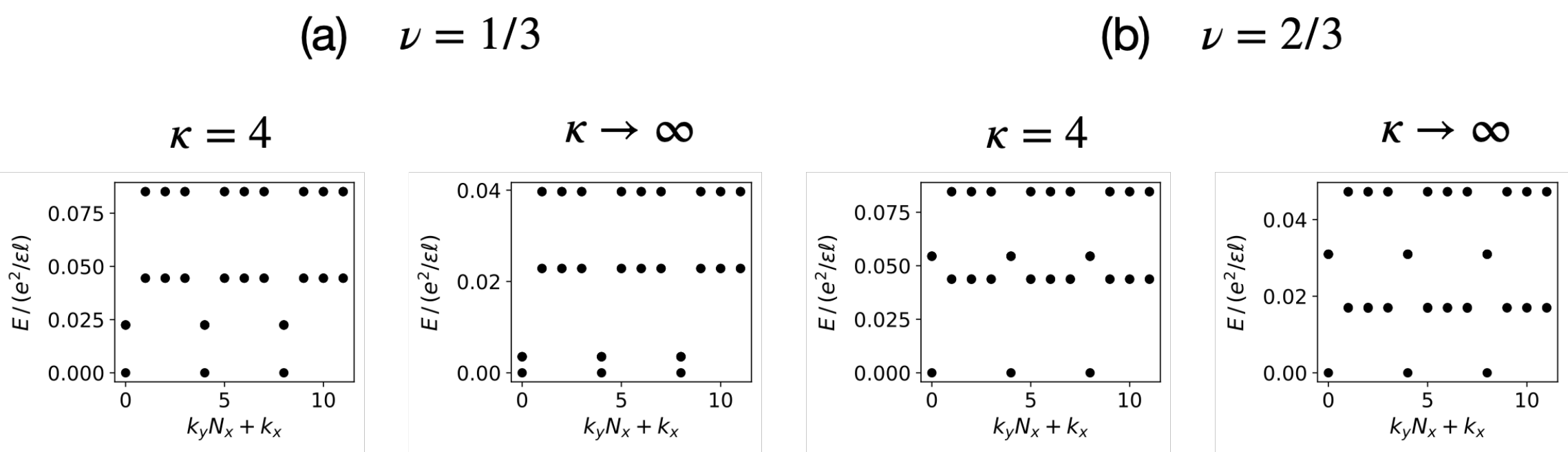}
    
    \caption{Comparison between many-body spectra with band-mixing parameters $\kappa=4$ and $\kappa\to\infty$ for (a) $\nu=1/3$ and (b) $\nu=2/3$ for the case of Landau levels, namely $B_1=U_1=0$ in Eq. \eqref{eq:Single_Particle_Hamiltonian}.}
    \label{fig:Inf_kappa}
\end{figure*}

For completeness, in Fig. \ref{fig:Inf_kappa} we show the many-body spectra obtained in the limit $\hbar\omega_c\to 0$ when the $n=0$ and $n=1$ Landau levels are exactly degenerate, for both filling fractions $\nu=1/3$ and  $\nu=2/3$. We also show the many-body spectra for the largest value of $\kappa$ considered in the main text ($\kappa=4$). We see that the many-body spectrum evolves adiabatically from $\kappa=4$ up to $\kappa\to\infty$ at both filling fractions.

\begin{figure*}
    \centering
    \includegraphics[width=\textwidth]{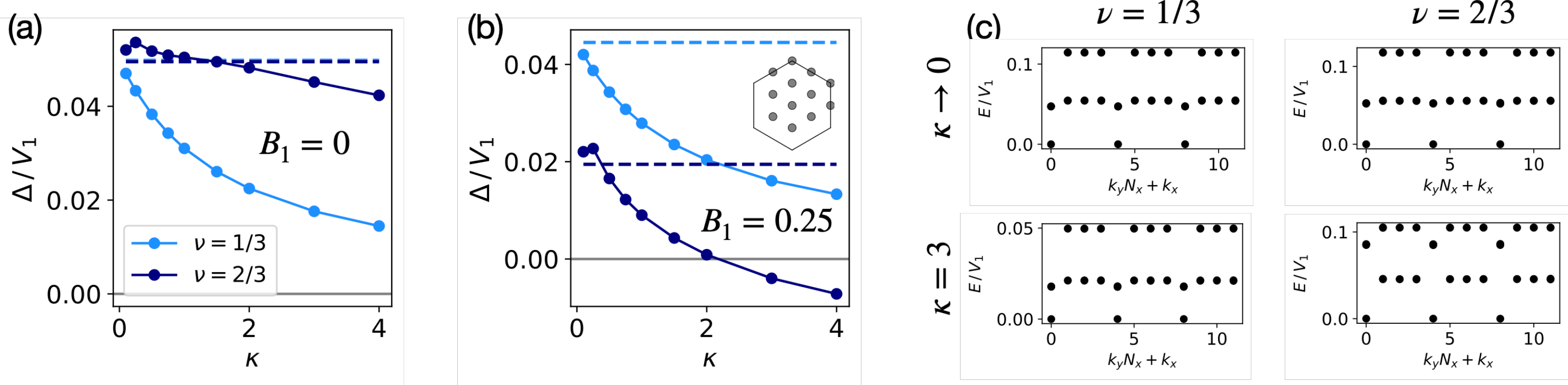}
    
    \caption{Evolution of the many-body gaps as a function of band-mixing for (a) Landau levels and (b) an AC band with $B_1=0.25$, using the Haldane pseudopotential interaction $V({\bm q})=-V_1 {\bm q^2}$. Dashed lines are one-band-projected calculations and lines with dots are two-band-projected calculations. (c) Typical many-body spectra for Landau levels at filling fractions $\nu=1/3$ and $\nu=2/3$ for two values of $\kappa$.}
    \label{fig:Haldane_Gaps}
\end{figure*}

In Fig. \ref{fig:Haldane_Gaps} we show the evolution of the many-body gaps at fillings $\nu=1/3$ and $\nu=2/3$ when the interaction term is given by the first Haldane pseudopotential. The results are in qualitative agreement with those in the main text obtained using the Coulomb interaction. For the case of Landau levels, Fig. \ref{fig:Haldane_Gaps}(a), the one-band-projected gaps (dashed lines) coincide, as expected. The two-band-projected gap at $\nu=1/3$ decreases monotonically, and the two-band-projected gap at $\nu=2/3$ shows a first increasing and then decreasing behavior, although the level crossing happens at a very small value of $\kappa$. This can be explained by the fact that the magnetoroton dispersion is almost flat for the Haldane pseudopotential interaction \cite{Repellin_MRM}, as can be seen in the many-body spectra shown in Fig. \ref{fig:Haldane_Gaps}(c). Results for an AC band with $B_1=0.25$ are shown Fig. \ref{fig:Haldane_Gaps}(b). The particle-hole symmetry is broken at the single-band-projected level due to the non-uniform quantum geometry of the band. We again observe the non-monotonic behavior of the two-band-projected gap at $\nu=2/3$. A recent single-band ED study on AC bands \cite{Guerci_AC_SC} finds a superconducting ground state at $\nu=2/3$ for large enough values of $B_1$. The effect of remote bands on the stability of that superconducting state is an interesting direction for future research.

\begin{figure*}
    \centering
    \includegraphics[width=\textwidth]{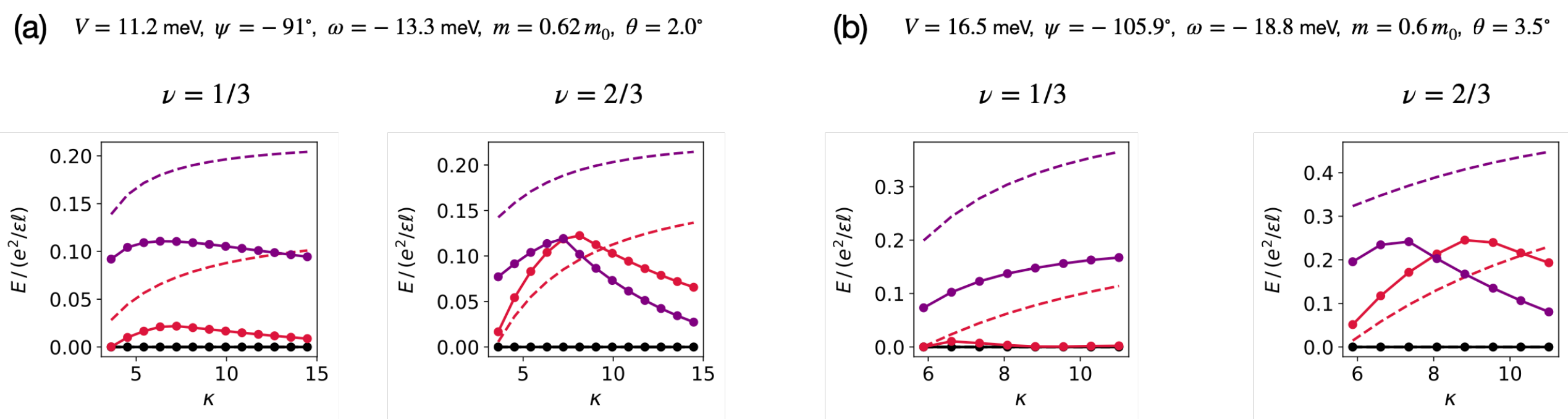}
    
    \caption{Evolution of the low-energy many-body states as a function of $\kappa$ at $\nu=1/3$ and $\nu=2/3$ for the two continuum models of tMoTe$_2$ studied here. Dashed lines correspond to single-band-projected calculations and lines with dots to two-band-projected calculations. The color convention is the same as in the main text. A level crossing can be observed at $\nu=2/3$ in both models.}
    \label{fig:LevelCrossings_MoTe2}
\end{figure*}

Finally, in Fig. \ref{fig:LevelCrossings_MoTe2} we show the evolution of the low-lying many-body energies as a function of the mixing parameter $\kappa$ for the two continuum models of tMoTe$_2$ studied in the main text. The behavior is in qualitative agreement with the LL, AC and adiabatic models. For two-band-projected calculations the lowest excited state always remains in the same momentum sector at $\nu=1/3$, while there is a level crossing between the two lowest excited states, which belong to different momentum sectors, at $\nu=2/3$. 

\subsection{Spherical geometry}
In this Appendix, we show results obtained in the spherical geometry. In this setup, a magnetic monopole is placed at the center of the sphere \cite{Haldane83}, such that the number of magnetic flux quanta is $2Q = \nu^{-1}N - S$ and the number of available orbitals is $N_\phi = 2Q+1$. Here, $S$ stands for the Wen-Zee shift \cite{WenZee}, which for states in the Jain sequence at $\nu = n/(2pn\pm 1)$ takes the value $S = 2p \pm n$.
The single-particle Hamiltonian is that of Eq.~(\ref{eq:Single_Particle_Hamiltonian}), restricted to the case of Landau levels by taking $B_1=U_1=0$.

\begin{figure*}[ht]
    \centering
    \includegraphics[width=\textwidth]{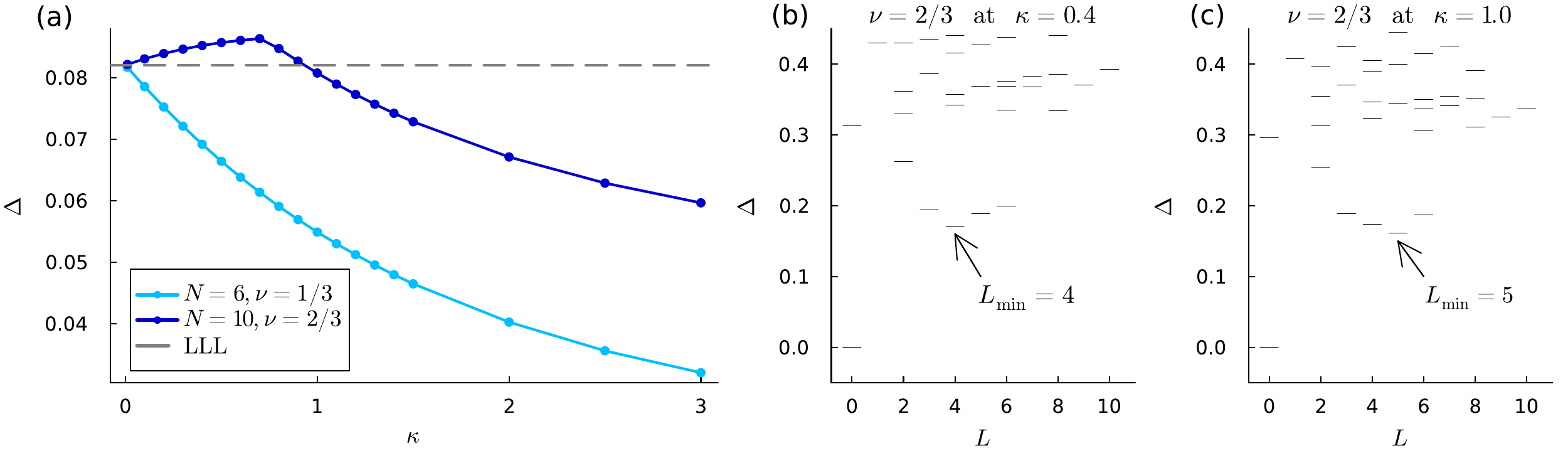}
    
    \caption{Exact diagonalization results on the sphere. The two conjugate filling factors are studied for the same number of flux quanta, $2Q=15$. (a) Many-body gap as a function of the band-mixing parameter $\kappa$. Similarly to the torus geometry, at $\nu = 1/3$, the gap decays monotonically, while at $\nu = 2/3$, the gap increases until $\kappa \approx 0.7$ and then decreases. (b)-(c) The low-energy spectrum at $\nu=2/3$ for values of $\kappa$ above and below the critical value $\kappa \approx 0.7$, resolved by total angular momentum $L$. The change in the behavior of the many-body gap near $\kappa \approx 0.7$ corresponds to a change in the location of the magnetoroton minimum from $L=4$ ($\kappa < 0.7$) to $L=5$ ($\kappa > 0.7$).}
    \label{fig:sphere}
\end{figure*}

The effect of band mixing on the many-body gap at filling fractions $\nu=1/3$ and $\nu=2/3$ is displayed in Fig.~\ref{fig:sphere}, showing remarkable agreement with the behavior of the many-body gap on the torus shown in Fig.~\ref{fig:LLL_MBGap} in the main text. 
Specifically, the kink in the many-body gap around $\kappa \approx 0.7$ (Fig.~\ref{fig:sphere}(a)) corresponds to a shift in the magnetoroton minimum from $L_\text{min} = 4$ to $L_\text{min} = 5$ (Fig.~\ref{fig:sphere}(b)-(c)). The increase in angular momentum of the minimum is consistent with the electron crystal, which has a smaller unit cell at $\nu=2/3$, becoming favorable at larger values of the band-mixing parameter $\kappa$. 

The change in the location of the magnetoroton minimum on the sphere can be understood as follows. Competing crystalline states are dictated by Thomson configurations \cite{Thomson1904, Archer13} -- in the thermodynamic limit they approach the triangular lattice, while in small systems there can be several competitive configurations, usually possessing large rotational symmetry. Numerically, these symmetry-broken states manifest through a ``tower of states", with low-lying states present in the angular momentum sectors allowed by the remaining symmetry subgroup of $SO(3)$. 
Let us now specialize to $2Q = 15$, corresponding to the calculations in Fig.~\ref{fig:sphere}. The electron crystal at $\nu=1/3$ and hole crystal at $\nu=2/3$ both form octahedra ($N=6$ vertices) that are very stable due to the large rotation symmetry group. The first allowed multipole moments in the  corresponding tower of states are $L=4,6,8\dots$, which is why the minimum occurs at $L=4$ for $\kappa  \lesssim  0.7$ in Fig.~\ref{fig:sphere}. After the level crossing at $\nu=2/3$, the roton minimum moves to $L=5$, which is forbidden for the hole crystal but allowed for the electron crystal, i.e. the $N=10$ Thomson configurations. 

\end{document}